\documentclass[journal=langd5,manuscript=article]{achemso}

\usepackage[version=3]{mhchem} 
\usepackage{times,epsfig,amsmath,amssymb,amsbsy,subeqnarray,psfrag}

\author{J.D. Sherwood}
\affiliation{Department of Applied Mathematics and Theoretical Physics, University of Cambridge, Wilberforce Road, Cambridge, CB3 0WA, UK}
\email{jds60@cam.ac.uk}
\author{M. Mao}
\author{S. Ghosal}
\affiliation{Department of Mechanical Engineering, Northwestern University, 2145 Sheridan Rd, Evanston, IL 60208, USA}
\altaffiliation{Department of
Engineering Sciences and Applied Mathematics, Northwestern University, 2145 Sheridan Rd, Evanston, IL 60208, USA}

\title[Electroosmosis in a finite cylindrical pore]
{Electroosmosis in a finite cylindrical pore: simple models of end effects}

\begin{document}

\begin{tocentry}

\begin{center}
\includegraphics[width=0.73\textwidth]{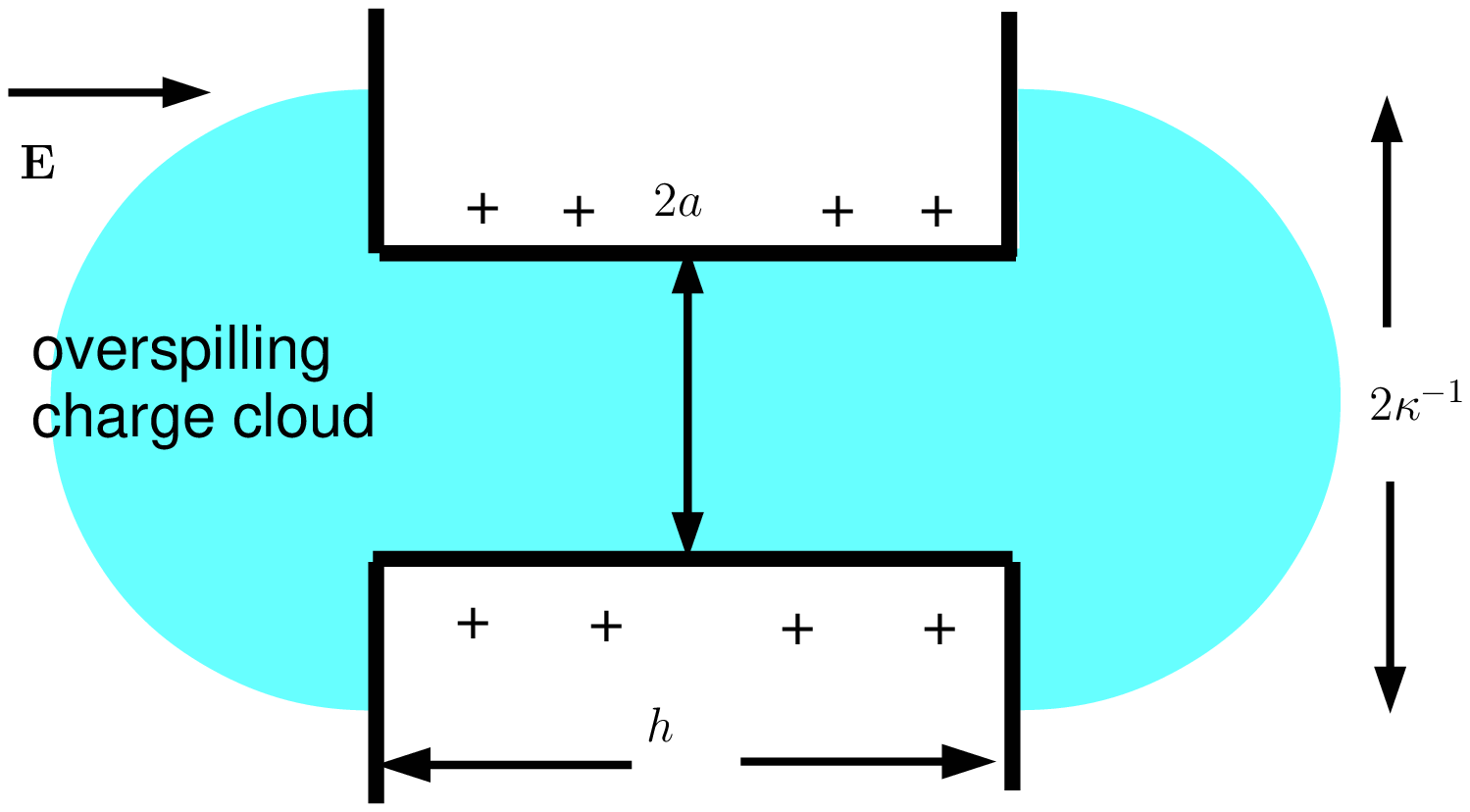}
\end{center}

\end{tocentry}

\begin{abstract}
A theoretical model of
electroosmosis through a circular pore of radius $a$ that
traverses a membrane of thickness $h$ is investigated.
Both the cylindrical surface of the pore and the outer surfaces
of the membrane are charged.
When $h\gg a$ end effects are negligible: results
of full numerical computations of
electroosmosis in an infinite pore agree with theory.
When $h=0$, end effects dominate, and computations again agree with analysis.
For intermediate values of $h/a$, an approximate analysis that combines
these two limiting cases captures the main features of computational
results when the Debye length $\kappa^{-1}$ is small compared with the pore
radius $a$. However, the approximate analysis
fails when $\kappa^{-1}\gg a$, when the charge
cloud due to the charged cylindrical walls of the pore spills out of
the ends of the pore, and the electroosmotic
flow is reduced. When this spilling out is included in the analysis,
agreement with computation is restored.

\end{abstract}



\bibliography{achemso-demo}

\newbox\grapha
\newbox\graphf
\newbox\graphg
\newbox\graphh

\setbox\grapha=\hbox{\hskip 2pt
\vrule height3pt depth-2pt width 40pt
\hskip 2pt}

\setbox\graphf=\hbox{\hskip 2pt
\vrule height3pt depth-2pt width 12pt
\hskip 4pt
\vrule height3pt depth-2pt width 12pt
\hskip 4pt
\vrule height3pt depth-2pt width 12pt
\hskip 4pt
\vrule height3pt depth-2pt width 12pt
\hskip 4pt}

\setbox\graphh=\hbox{\hskip 2pt
\vrule height3pt depth-2pt width 9pt
\hskip 4pt
\vrule height3pt depth-2pt width 2pt
\hskip 4pt
\vrule height3pt depth-2pt width 9pt
\hskip 4pt
\vrule height3pt depth-2pt width 2pt
\hskip 4pt
\vrule height3pt depth-2pt width 9pt
\hskip 4pt}

\setbox\graphg=\hbox{\hskip 2pt
\vrule height3pt depth-2pt width 2pt
\hskip 4pt
\vrule height3pt depth-2pt width 2pt
\hskip 4pt
\vrule height3pt depth-2pt width 2pt
\hskip 4pt
\vrule height3pt depth-2pt width 2pt
\hskip 4pt
\vrule height3pt depth-2pt width 2pt
\hskip 4pt
\vrule height3pt depth-2pt width 2pt
\hskip 4pt
\vrule height3pt depth-2pt width 2pt
\hskip 4pt}

\section{1. Introduction}

Electroosmosis in a circular cylindrical pore of finite length $h$ differs from
that in an infinitely long pore due to end effects. If the cylinder length
$h=0$, the pore consists of a hole in a charged membrane of zero thickness, and
electroosmosis can be considered to be entirely due to end
effects. This case was considered by us previously \cite{mao14}.
When the cylindrical pore is
infinitely long, end effects are negligible, and the computation of
the electroosmotic volumetric flow rate $Q$ is 
straightforward\hbox{\cite{rice1965,gross1968}}. 
Here we are interested in intermediate values of $h$. 
Similar results are also available in planar infinitely long channels for 
arbitrary Debye length and wall charge \cite{baldessari2008a,baldessari2008b}.
A related problem of interest is the flow generated by an electric field 
along a flat surface with a step change in surface charge density \cite{yariv2004,khair2008}.

Full numerical computation of the Poisson-Nernst-Planck (PNP) equations for
ionic motion is of course possible, and some typical results were
reported by Mao et al. \cite{mao14}.
Such numerical computations however do not identify the mechanisms underlying the qualitative 
features of the physical system.
Here we discuss how simple models, based
upon continuity of electric current and volumetric flow rate, can be
combined in order to estimate end effects for pore lengths $h>0$.
We assume that the zeta potential on the surface of the membrane is small,
so that the Poisson-Boltzmann equation governing the equilibrium charge
cloud can be linearized, and the electroosmotic
velocity can be determined by an analysis equivalent to that of
Henry \cite{henry31} for electrophoresis, i.e.
fluid motion is generated by the effect of the applied electric field
acting on the equilibrium charge cloud (which is not deformed either by the
applied electric field or by fluid motion).
In this limit the electroosmotic volumetric flow rate $Q$ through the hole in
the membrane can be determined
by means of the reciprocal theorem \cite{mao14}.

\begin{figure}
\begin{center}
\includegraphics[width=0.7\textwidth]{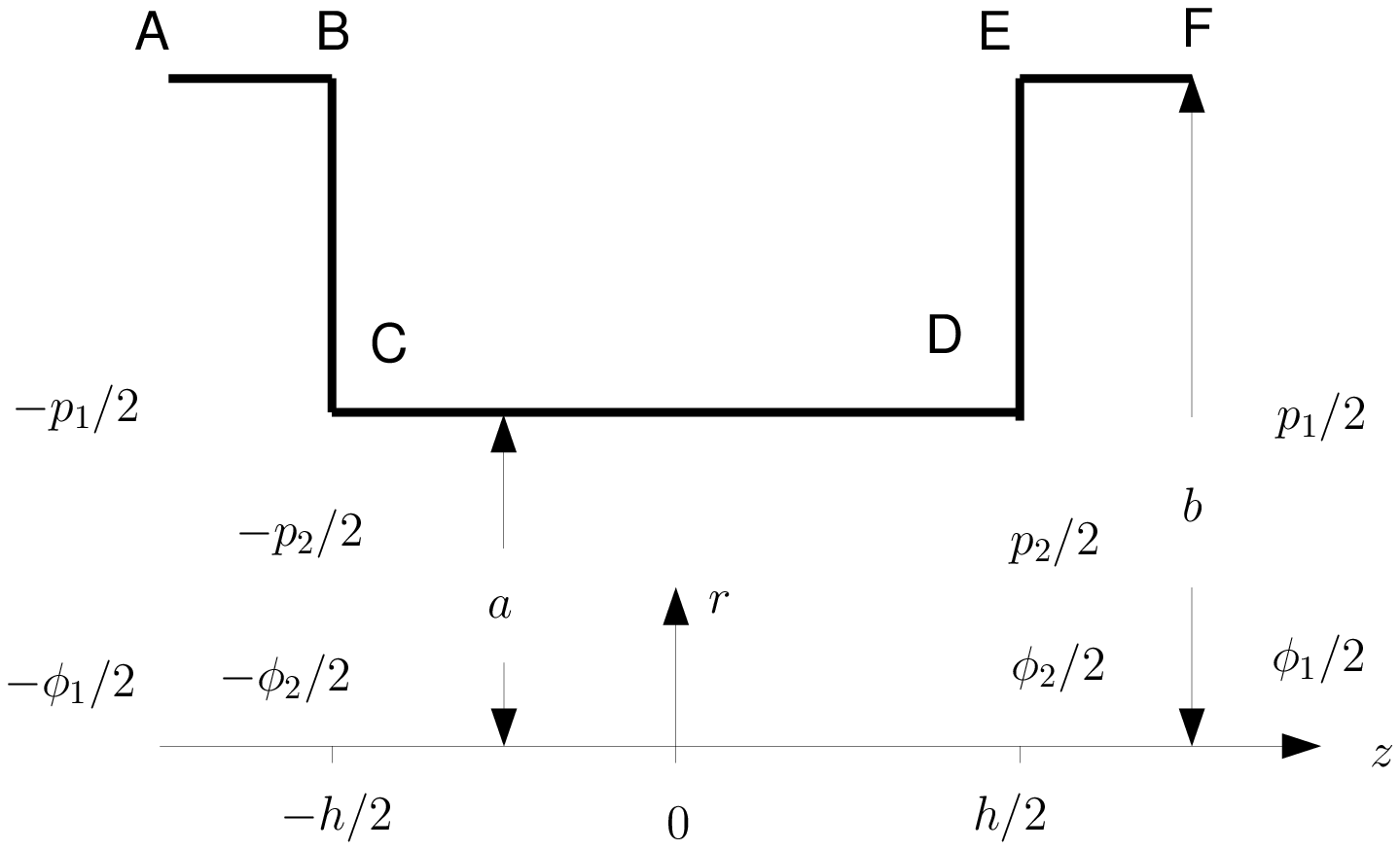}
\caption{\label{Fig:schematic}
The cylindrical pore CD, of length $h$ and
radius $a$ with surface charge density $\sigma_c$, passing through the
membrane with surface charge density $\sigma_m$ on the two surfaces BC
and DE. The reservoirs on either side of the membrane are large
($b\gg a$). The pore and reservoirs are axisymmetric about the $z$ axis.}
\end{center}
\end{figure}

\ref{Fig:schematic} shows the axisymmetric geometry that we are
considering. The cylindrical pore CD has radius $a$ and length $h$.
The cylindrical surface CD of the pore has
surface charge density $\sigma_c$, and the
membrane surfaces BC and DE have surface charge density $\sigma_m$.
An electrical potential difference is applied betwen the fluid reservoirs
at either side of the membrane, and
electroosmotic flow is generated by the resulting electric field
acting on the charge cloud adjacent to the charged surfaces.
The analysis of Mao et al.\cite{mao14} assumed that the external reservoirs
on either side of the pore were unbounded, with radius $b=\infty$.
For numerical work,
the external reservoirs were bounded by uncharged cylinders of
radius $b\gg a$,  sufficiently large that numerical
results when $h=0$
differed little from the analytic results for $h=0$ and $b$
infinite.  There have been many studies in which flow is
generated in cylinders of different dimensions, connected either
in series \cite{biscombe12}
or in networks intended to represent porous media \cite{jin91}. Here,
however, we are interested in the effect of the surfaces BC and DE of
the membrane on electroosmotic flow within the cylindrical pore,
and any boundaries AB, EF of the external reservoirs are so far away that
they can be neglected.

In section 2 we set up the approximate analysis of end effects,
and compare results to those obtained from full numerical
computations. The analysis is presented from first principles, but can
alternatively be set within the framework of the reciprocal theorem,
as explained in section 2.6. The agreement between the approximate
analysis and full computation is in general good, except for large Debye
lengths $\kappa^{-1}\gg a$. In section 3 we consider this case in more
detail, in order to evaluate how much of the charge cloud due to the
charged walls of the cylindrical pore lies within the pore and how
much spills out beyond the ends of the pore. When this overspill is taken into
account, the agreement between the computations and the approximate
model is improved.

\section{2. Composite electroosmotic coefficient}

\subsection{2.1 The pore geometry}

The axisymmetric geometry that we are considering is shown in \ref{Fig:schematic}.
We use cylindrical polar coordinates $(r,z)$, with the $z$ axis along
the axis of symmetry and $z=0$ at the midpoint of the cylindrical
pore, the ends of which are at $z=\pm h/2$. When $h=0$ we shall also use
oblate spherical coordinates $(\xi,\eta)$, with
\begin{equation}
z=a\sinh\xi\cos\eta\quad,\quad r=a\cosh\xi\sin\eta,
\end{equation}
where $-\infty<\xi<\infty$ and $0\le \eta<\pi/2$.

The cylindrical pore and the reservoirs at either end are filled with liquid
with electrical conductivity $\Sigma$ and viscosity $\mu$.
The wall CD of the cylindrical pore is charged, with uniform
surface charge density $\sigma_c$, and the surface charge density
over the membrane surfaces BC, DE, is $\sigma_m$.
We assume that the reservoir boundaries AB, EF are uncharged and at infinity.
We shall occasionally refer to the surface potential $\zeta$, which
will not in general be uniform, but which is required to be small, with
$\zeta\ll kT/e$,
where $e$ is the elementary charge and $kT$ the Boltzmann
temperature.  
The electrical potential $\phi_0$ within the equilibrium charge cloud 
therefore satisfies
the linearized Poisson-Boltzmann equation, so that
\begin{equation}
\nabla^2\phi_0=\kappa^2\phi_0,
\label{linear_poisson_boltzmann_eqn}
\end{equation}
and the charge density in the equilibrium charge cloud is
\begin{equation}
\rho_0=-\epsilon\kappa^2\phi_0.
\end{equation}

\subsection{2.2 The applied electric field}

The applied electric field is
$\mathbf{E}=-\nabla\chi$,
where the potential $\chi$ satisfies the Laplace equation
\begin{equation}
\nabla^2\chi=0,
\end{equation}
with gradient
\begin{equation}
\mathbf{n}.\nabla\chi=0
\end{equation}
normal to the walls of the membrane and of the cylindrical pore. 
In $z>0$, the electric potential far from the membrane
is $\chi=\phi_1/2$, and the potential
far from the membrane in $z<0$ is
$\chi=-\phi_1/2$.

When the membrane thickness $h=0$, the potential can be expressed
explicitly as \cite{morse53}
\begin{equation}
\chi=\frac{\phi_1}{2}\left\lbrack 1-\frac{2}{\pi}
\tan^{-1}\left(\frac{1}{\sinh\xi}\right)\right\rbrack
=\tilde\chi_m(r,z)\phi_1.
\label{chi_m}
\end{equation}
On the plane of the membrane, within the circular opening,
\begin{equation}
\tilde\chi_m=0,\hskip 20pt z=0,\ r<a,\ h=0.
\label{chi_m_z0}
\end{equation}
The liquid within the pore has electrical conductivity $\Sigma$;
we have assumed that surface charge density (and hence the density of charge in
the cloud of counter ions) is small, so that surface conductivity
may be neglected.
The total electric current $I_m$ flowing through the hole in the membrane
is therefore
\begin{equation}
I_m=-\frac{\phi_1}{R_m}\quad,\quad R_m=\frac{1}{2a\Sigma}.
\label{Im_h0}
\end{equation}

If $h>0$, we assume that the potential within the cylindrical pore
varies linearly and 
approximate the potential within the pore as
\begin{equation}
\chi=\tilde\chi_c\phi_2=\frac{z}{h}\phi_2,\hskip 20pt r<a,\ |z|<h/2,
\label{approx_chi_c}
\end{equation}
as would be expected in the absence of any end effects.
The potential in $z>h/2$ is approximated by that outside
a membrane (with a hole) of zero thickness:
\begin{equation}
\chi=\frac{\phi_2}{2}+(\phi_1-\phi_2)\tilde\chi_m(r,z-h/2),
\label{approx_chi_m}
\end{equation}
with $\chi(r,z)=-\chi(r,-z)$. This approximation (\ref{approx_chi_c})
and (\ref{approx_chi_m}) is continuous at
$z=\pm h/2$ where the potential is assumed to be
$\phi_2$/2 across the entire width
of the opening (by (\ref{chi_m_z0})). The as yet unspecified
potential $\phi_2$ is determined by requiring continuity of the
electrical current at $z=\pm h/2$.
The current $I_c$ through the cylindrical pore is
\begin{equation}
I_c=-\frac{\phi_2}{R_c}\quad,\quad R_c=\frac{h}{\pi a^2\Sigma},
\label{I_c}
\end{equation}
and the electrical current through the reservoir
in $z>h/2$ is, by (\ref{Im_h0}),
\begin{equation}
I_m=- \frac{\phi_1-\phi_2}{R_m}.
\label{I_m}
\end{equation}
Equating $I_c$ (\ref{I_c}) and  $I_m$ (\ref{I_m}), we find
\begin{equation}
\phi_2=\frac{R_c\phi_1}{R_m+R_c}.
\label{phi_2}
\end{equation}

\subsection{2.3 Electroosmosis through an infinite cylindrical pore}

We assume throughout this paper that the perturbation of
the equilibrium charge cloud by the applied electric field and by
fluid motion is negligibly small.
The force acting on the ions in the
charge cloud due to the applied electric field $-\nabla\chi$ is
therefore $-\rho_0\nabla\chi$.

The equilibrium potential within an infinite cylindrical pore is
\begin{equation}
\phi_0=\zeta_c\frac{I_0(\kappa r)}{I_0(\kappa a)}
=\frac{\sigma_c}{\epsilon\kappa}\frac{I_0(\kappa r)}{I_1(\kappa a)}.
\label{phi0_cylinder}
\end{equation}
In the absence of any end effects, if the electric field
$E_0=-\phi_2/h$ is applied along the length of the cylindrical pore,
the fluid velocity is \cite{levine1975}
\begin{equation}
u=\frac{\epsilon\phi_2}{\mu h}(\zeta_c-\phi_0),
\end{equation}
and the total electroosmotic volumetric flow rate is
\cite{rice1965}
\begin{equation}
Q_{ce}=\frac{2\pi \sigma_ca^3}{\mu h}\left\lbrack
\frac{1}{2\kappa a}\frac{I_0(\kappa a)}{I_1(\kappa  a)}
-\frac{1}{(\kappa a)^2}\right\rbrack\phi_2=H_c\phi_2,
\label{Hc_defn}
\end{equation}
where the electroosmotic coefficient
\begin{subeqnarray}
H_c&\sim& \frac{\pi\sigma_c a^2}{\mu h\kappa}
=\frac{\pi a^2\zeta_c}{\mu h\epsilon},\hskip 20pt a\kappa\gg 1,
\\
&\sim&\frac{\pi\sigma_ca^3}{4\mu h},\hskip 20pt a\kappa\ll 1.
\slabel{H_c_kappa_small}
\end{subeqnarray}

\subsection{2.4 Electroosmosis through a membrane ($h=0$)}

It was shown by Mao et al. \cite{mao14} that if the equilibrium charge density
is $\rho_0$, the imposed electric field is
$\mathbf E=-\nabla\chi$ and the fluid velocity generated by a pressure
difference $p_1$ across a pore (of arbitrary geometry) is
\begin{equation}
\mathbf{u}=p_1\mathbf{G},
\end{equation}
then the reciprocal theorem \cite{happel73} for Stokes flows
can be used to
show that electroosmotically generated volumetric flow rate through the pore is
\begin{equation}
Q=-\int_V\rho_0\mathbf{G}.\nabla\chi\,\text{d}V,
\label{reciprocal_integral}
\end{equation}
where the integral is over all the fluid.

The fluid velocity generated by the pressure difference $p_1$
across a circular hole in a membrane of zero thickness is
\begin{equation}
\mathbf{u}=p_1\mathbf{G}^m.
\label{u_membrane}
\end{equation}
An explicit expression for $\mathbf{G}^m(r,z)$ is available
\cite{happel73,mao14}, and the potential $\chi$ is given by (\ref{chi_m}).
The charge density in the equilibrium charge cloud around a
membrane of zero thickness is \cite{mao14}
\begin{equation}
\rho_0 = \sigma_m\kappa^2 a \left[  \int_0^\infty
\frac{J_1(as)J_0(rs)}{(\kappa^2+s^2)^{1/2}}
e^{-(\kappa^2+s^2)^{1/2} z}\,\text{d}s - \frac{e^{-\kappa z}}{\kappa a} \right],
\label{eq:rho0}
\end{equation}
which consists of the charge density adjacent to a uniform charged surface,
from which has been subtracted the charge density around a
uniformly charged disk. The integral (\ref{reciprocal_integral})
can be evaluated numerically
\cite{mao14},
and the electroosmotic flow rate
through a hole in a membrane of zero thickness can be expressed in the form
\begin{equation}
Q_{me}=H_m\phi_1,
\label{Hm_defn}
\end{equation}
where
\begin{equation}
H_{m}\sim a\kappa H_0,\hskip 20pt a\kappa\ll 1,
\label{H_m_akappa_small}
\end{equation}
with
\begin{equation}
H_0=\frac{a^2\sigma_m}{3\mu}.
\label{H0_defn}
\end{equation}

\begin{figure}
\begin{center}
\includegraphics[width=0.46\textwidth]{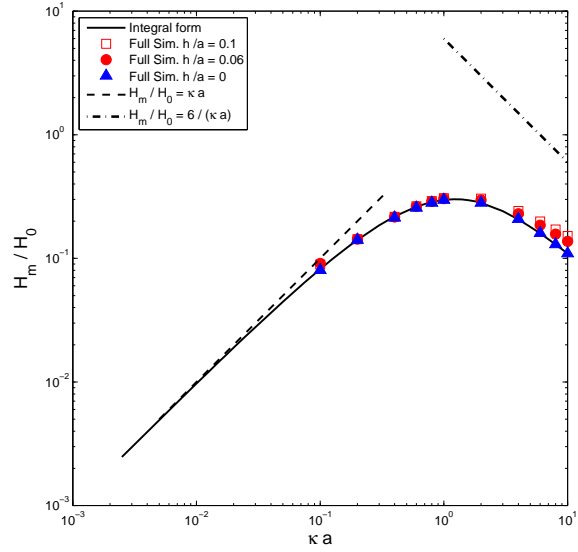}
\end{center}
\caption{\label{Fig:H_m_log_log}
The electroosmotic coefficient $H_{m}$, 
scaled by $H_0$ (\ref{H0_defn}), for a membrane of thickness $h=0$,
as a function of $a\kappa$. \copy\grapha analytic result
(\ref{reciprocal_integral});
\copy\graphf asymptote (\ref{H_m_akappa_small}) for $a\kappa\ll 1$;
triangles: full PNP numerical computation ($h=0$).
The dot-dashed line \copy\graphh\ shows $H_{m}/H_0=6/(a\kappa)$
with the expected slope for large $a\kappa$.
Squares and circles show electroosmotic coefficients $H/H_0$
for non-zero membrane thickness $h>0$, computed 
by numerical integration of the full PNP equations:
solid circles  $h/a=0.06$; open squares $h/a=0.1$.
}
\end{figure}

\ref{Fig:H_m_log_log}
shows a log-log plot of results for $H_m/H_0$ obtained by
Mao et al. \cite{mao14}. The continuous line shows the analytic result
(\ref{reciprocal_integral})
obtained via the reciprocal theorem, and
the asymptote (\ref{H_m_akappa_small}) for $a\kappa\ll
1$ is indicated. 

The membrane has zero thickness, so that there is always a region near
the edge of the pore where the Debye length $\kappa^{-1}$ cannot be
considered small compared with $h$:
Smoluchowski's analysis for thin charge
clouds, which would predict $H=6H_0/(a\kappa)$ if $\zeta_m$ took the
uniform value $\epsilon\kappa\sigma_m$,
therefore cannot automatically be invoked when $a\kappa\gg 1$.
However, if we set up a local coordinate $s$ indicating distance from
the edge of the pore, both the electric potential $\chi$ (\ref{chi_m})
and the fluid velocity $\mathbf{G}^m$ (\ref{u_membrane})
vary as $s^{1/2}$ when $s\ll a$ (i.e. near the pore edge).
The charge cloud density $\rho_0$ decays over a lengthscale $\kappa^{-1}$,
and only counter-ions of membrane surface charge within a distance $\kappa^{-1}$
from the edge contribute to $\rho_0$ within the hole. The contribution
of the edge to the integral (\ref{reciprocal_integral})
is therefore $O((a\kappa)^{-1})$,
as was similarly found for the electrophoretic velocity of a charged disk \cite{sherwood95}.
We therefore expect
$H_m\sim KH_0/(a\kappa)$ when $a\kappa\gg 1$. The data in
\ref{Fig:H_m_log_log} do not extend to sufficiently high
values of $a\kappa$ to allow us to estimate $K$ with
any accuracy, and for the figure we simply indicate the line
$K=6$ suggested by the Smoluchowski analysis.
Clearly the asymptote corresponds to a value $K<6$.
A similar reduction in the broadside electrophoretic velocity of a disk
below the value predicted by Smoluchowski was
noted by Sherwood \& Stone.\cite{sherwood95}
Individual points in \ref{Fig:H_m_log_log}
indicate results obtained from full numerical solutions of
the Poisson-Nernst-Planck equations in a symmetric electrolyte
at low applied potential and low surface charge. 
In the computations, the length of
the reservoirs in the $z$ direction was equal to their
radius $b$, with $b=\max(10a,10\kappa^{-1})$.
Other details of the computations are
reported in section~4.

\subsection{2.5 Composite electroosmotic coefficient $H_{\rm comp}$}

When $h>0$ it is natural to suppose that the electric field
ouside the membrane pumps fluid towards the cylindrical pore at a rate
\begin{equation}
Q_{me}\approx H_m(\phi_1-\phi_2),
\label{q_me}
\end{equation}
and the electric field within the cylindrical pore
pumps fluid through the pore at a rate
\begin{equation}
Q_{ce}\approx H_c\phi_2.
\label{q_ce}
\end{equation}
However, in general, $Q_{me}$ (\ref{q_me}) and
$Q_{ce}$ (\ref{q_ce}) differ, and a pressure $\pm p_2/2$ builds
up at $z=\pm h/2$ (i.e. at the entrance and exit to the cylindrical pore)
in order to ensure that the volumetric flow rate is
continuous. We now determine this pressure $p_2$.

Consider a membrane of zero thickness ($h=0$), with pressure $p=p_1/2$
(above the reference ambient pressure)
at infinity on the side $z>0$, and with $p=-p_1/2$ at infinity on
the other side. The pressure within the hole in the membrane is
\begin{equation}
p=0,\quad z=0,\ r<a,\ h=0.
\end{equation}
The fluid velocity generated by the pressure difference $p_1$
across the membrane is
$\mathbf{u}=p_1\mathbf{G}^m$ (\ref{u_membrane}),
and the corresponding volumetric flow rate is \cite{happel73} 
\begin{equation}
Q_{mh}=G_mp_1,\quad G_m=-\frac{a^3}{3\mu}.
\end{equation}

If $h>0$ we approximate the pressure field in the fluid in much the same way as
we approximated the electrical potential within the fluid:
we patch a linearly varying pressure $p(z)$ within the
cylindrical pore to the pressure field outside a membrane of zero thickness,
and we take the pressure over the two ends $z=\pm h/2$ of the cylindrical
pore to be $\pm p_2/2$.
Thus the pressure within the pore is approximated as
\begin{equation}
p=\frac{p_2}{h}z,\quad r<a,\ |z|<h/2,
\end{equation}
the fluid velocity within the pore is
\begin{equation}
\mathbf{u}=p_2\mathbf{G}^c,
\label{u_c}
\end{equation}
and the volumetric flow rate within the pore is
\begin{equation}
Q_{ch}=G_cp_2,\quad G_c=-\frac{\pi a^4}{8h\mu}.
\label{G_c}
\end{equation}
Outside the cylindrical pore, the fluid velocity is assumed now to be
\begin{equation}
\mathbf{u}=(p_1-p_2)\mathbf{G}^m(r,z-h/2),\quad z>h/2,
\label{u_m}
\end{equation}
with $u_r(r,z)=-u_r(r,-z)$ and $u_z(r,z)=u_z(r,-z)$. 
The volumetric flow rate outside the membrane is now
\begin{equation}
Q_{mh}=G_m(p_1-p_2),\quad G_m=-\frac{a^3}{3\mu}.
\label{G_m}
\end{equation}
We have
ensured that the pressure (but not the fluid
velocity nor the volumetric flow rate) is continuous across the ends
$z=\pm h/2$ of
the cylindrical pore.

When an electric field generates an electroosmotic velocity, the
volumetric flow rates within the cylindrical pore and outside the
membrane are identical if $p_2$ is such that $Q_{mh}+Q_{me}=Q_{ch}+Q_{ce}$,
i.e. if
\begin{equation}
G_m(p_1-p_2)+H_m(\phi_1-\phi_2)=G_cp_2+H_c\phi_2.
\end{equation}
But the pressure at infinity is zero in the electroosmotic problem, so
$p_1=0$, and $\phi_2$ is given by (\ref{phi_2}).
Hence
\begin{equation}
p_2=
\frac{H_mR_m-H_cR_c}{(G_m+G_c)(R_m+R_c)}\phi_1,
\end{equation}
and the total electro-osmotic flow is
\begin{equation}
Q_E=Q_{me}+Q_{mh}=
\frac{(G_mR_cH_c+G_cH_mR_m)}{(R_m+R_c)(G_m+G_c)}\phi_1=H_\text{comp}\phi_1.
\label{flow_rate_lumped_parameter}
\end{equation}
An alternative derivation of this approximate composite
$H_\text{comp}$ (\ref{flow_rate_lumped_parameter}) is given in the next
section.

Inserting into (\ref{flow_rate_lumped_parameter})
the various estimates for $G_m$ (\ref{G_m}), $G_c$ (\ref{G_c}),
$R_m$ (\ref{Im_h0}) and $R_c$ (\ref{I_c}),
we obtain
\begin{equation}
H_\text{comp}=
\frac{\left(H_m+\frac{16 h^2}{3\pi^2a^2}H_c\right)}
{\left(1+\frac{2h}{\pi a}\right)
\left(1+\frac{8h}{3\pi a}\right)}.
\label{H_comp}
\end{equation}
For small $h/a$ the approximate composite $H_\text{comp}$
is
larger than $H_m$ if
\begin{equation}
\frac{H_c}{H_m}>\frac{7\pi a}{8h}.
\label{H_c_for_H_comp_increasing}
\end{equation}
Experimental arrangements often involve measurements at fixed current. 
Thus, it is helpful to define a coefficient that gives the electroosmotic flux per unit current. 
This quantity may be obtained readily from (\ref{I_c}), (\ref{phi_2}) and (\ref{flow_rate_lumped_parameter}):
\begin{equation} 
H_\text{comp}^{\prime} \equiv - \frac{Q_E}{I_c} = 
\frac{(G_mR_cH_c+G_cH_mR_m)}{(G_m+G_c)} = 
\frac{\left(H_m+\frac{16 h^2}{3\pi^2a^2}H_c\right)}
{\left(1+\frac{8h}{3\pi a}\right)}.
\end{equation} 

\subsection{2.6 Composite electroosmotic coefficient $H_{\rm comp}$
derived via the reciprocal theorem}

We now show that approximations to the electric potential $\chi$ and
pressure-driven velocity $\mathbf{G}$ within a pore of non-zero length $h>0$,
when inserted into the integral expression (\ref{reciprocal_integral})
for the electroosmotic
volume flux, lead to an approximate electroosmotic coefficient
identical to $H_\text{comp}$ (\ref{H_comp}) obtained in the previous section.

We have already shown that we may approximate the electric potential
by a composite potential (\ref{approx_chi_c}), (\ref{approx_chi_m}), of the form
\begin{subeqnarray}
\chi&=&\left(\frac{z}{h}\right)\frac{R_c\phi_1}{R_m+R_c},\hskip 80pt |z|<h/2,
\\
&=&\frac{R_c\phi_1}{2(R_m+R_c)}+\frac{R_m\phi_1}{R_m+R_c}
\tilde\chi_m(r,z-h/2),
\hskip 10pt z>h/2,
\\
&=&\chi(r,-z),\hskip 90pt z<0.
\label{chi_approx_composite}
\end{subeqnarray}
We now create a similar approximation for the fluid velocity for flow 
through a membrane of thickness $h$ subjected only to a pressure drop $p_1$ 
but no applied potential drop.
We suppose that in $z>h/2$ the fluid velocity is given by (\ref{u_m}),
corresponding to flow outside a membrane of zero thickness, and that
within the cylindrical pore the fluid velocity is given by
(\ref{u_c}). Continuity of the volumetric flow rates (\ref{G_c}), (\ref{G_m}) at the entrance to
the cylindrical pore requires that the pressure
$\pm p_2/2$ at the two ends of the pore satisfies
\begin{equation}
G_cp_2=G_m(p_1-p_2),
\end{equation}
so that
\begin{equation}
p_2=\frac{G_mp_1}{G_c+G_m}.
\end{equation}
Hence our approximation to the fluid velocity
is $\mathbf{u}=\mathbf{G}p_1$, with
\begin{subeqnarray}
\mathbf{G}&=&
\frac{G_m}{G_c+G_m}\mathbf{G}^c(r,z),\hskip 65pt |z|<h/2,
\\
&=&\frac{G_c}{G_c+G_m}\mathbf{G}^m(r,z-h/2),\hskip 30pt z>h/2.
\label{G_approx_composite}
\end{subeqnarray}

We now use the approximations (\ref{chi_approx_composite}) and
(\ref{G_approx_composite}) in the integral (\ref{reciprocal_integral}) in order to
compute the electroosmotic volumetric flow rate. But the integration
splits naturally into an integral over the cylindrical pore and an
integral over the regions outside the membrane. The integral over the
cylindrical pore is exactly the integral required to determine
the electroosmotic flow rate $H_c$ (\ref{Hc_defn}) in a cylinder, and the
integral outside the membrane is exactly that required to determine
$H_m$ (\ref{Hm_defn}). Hence the integral yields the composite
electroosmotic flow rate
\begin{equation}
H_\text{comp}= \frac{G_cR_mH_m}{(G_c+G_m)(R_m+R_c)}
+\frac{G_mR_cH_c}{(G_c+G_m)(R_m+R_c)}
=\frac{G_cR_mH_m+G_mR_cH_c}{(G_c+G_m)(R_m+R_c)},
\end{equation}
identical to (\ref{flow_rate_lumped_parameter}), obtained in section 2.5 by
elementary methods.

\subsection{2.7 Predictions of the composite electroosmotic coefficient}

\begin{figure}
\begin{center}
\includegraphics[width=\textwidth]{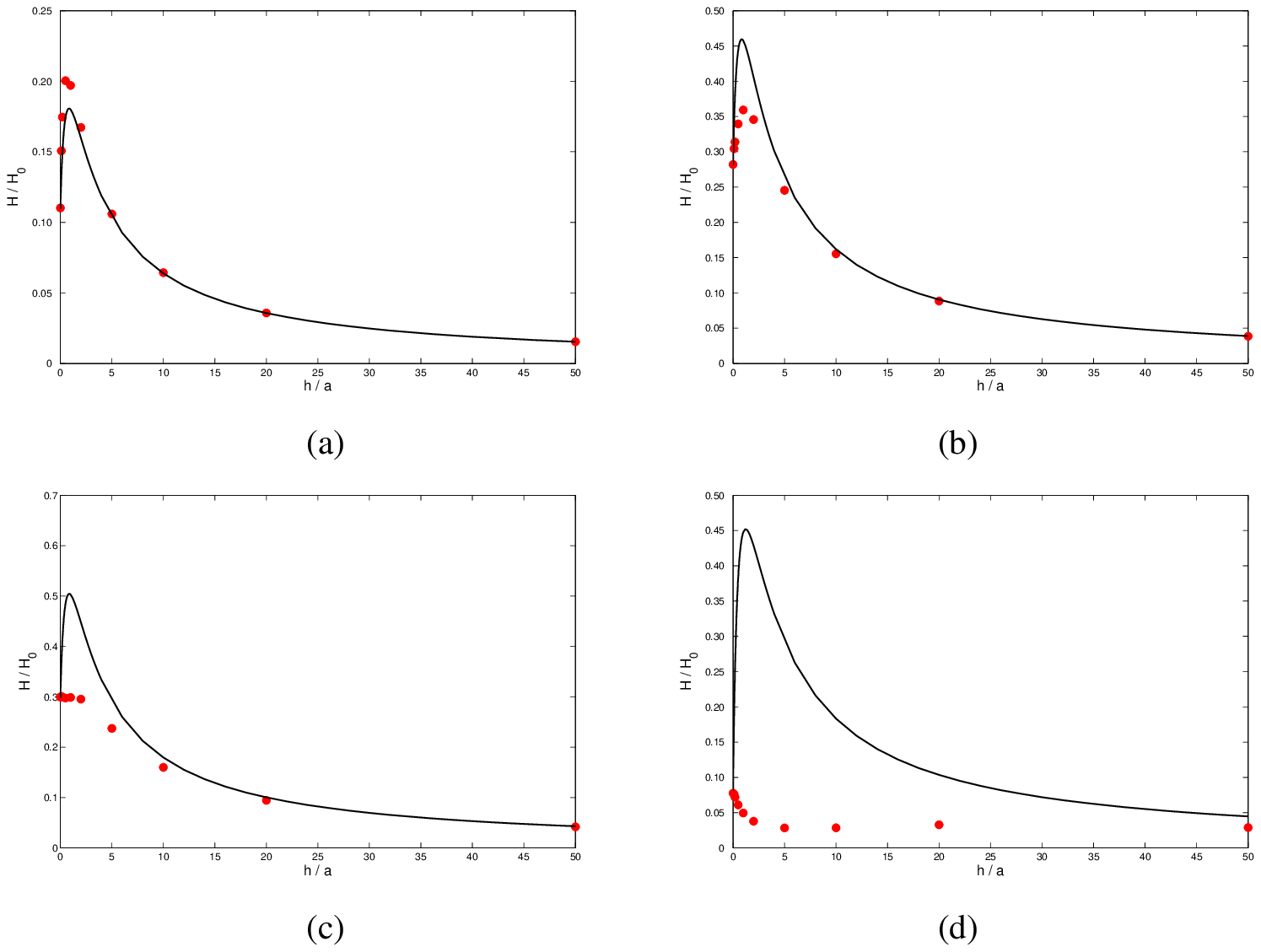}
\caption{\label{Fig:H_comp}
The electroosmotic coefficient $H$ scaled by $H_0$ (\ref{H0_defn}) for
$\sigma_m=\sigma_c$, as a function of $h/a$, for
(a) $a\kappa=10$, (b) $a\kappa=2$, (c) $a\kappa=1$,
 (d) $a\kappa=0.1$.
\copy\grapha $H_\text{comp}$ (\ref{H_comp});
solid circles: full PNP numerical computation.}
\end{center}
\end{figure}

\begin{figure}
\begin{center}
\includegraphics[width=0.6\textwidth]{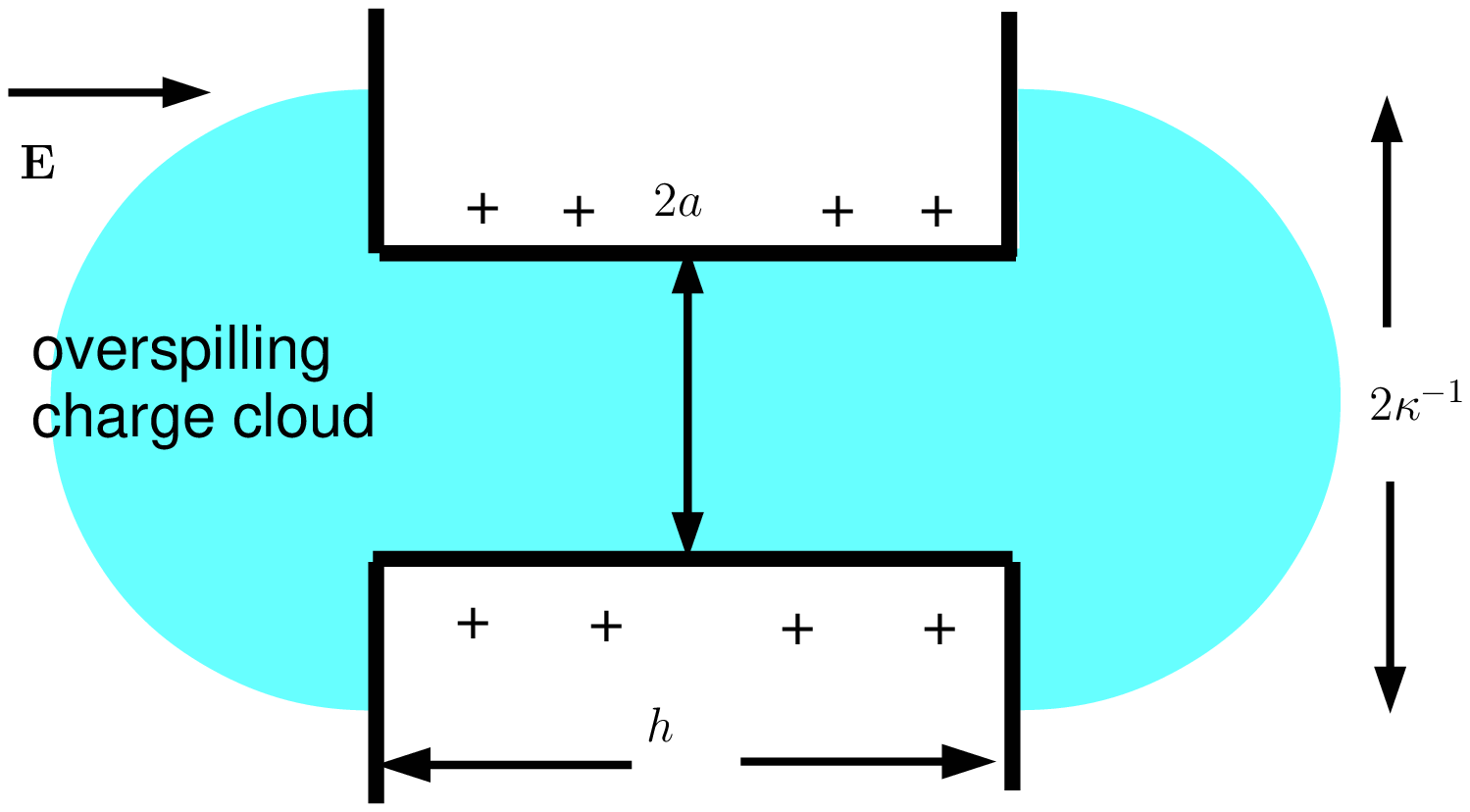}
\end{center}
\caption{\label{Fig:schematic_overspill}
When the Debye length $\kappa^{-1}$ is large compared with the pore radius $a$,
the cloud of counter ions associated with the charged cylindrical
wall of the pore spills out of the ends of the pore.
}
\end{figure}

\ref{Fig:H_comp} shows $H_\text{comp}$ (\ref{H_comp}) as a function of $h/a$, for four
different values of $a\kappa$, with $\sigma_m=\sigma_c$. Also shown are the results of full numerical
computations based on the
Poisson-Nernst-Planck equations \cite{mao14}. We see that for $a\kappa\ge 1$ the
approximate analysis captures the main features of the full numerical
results, and it is clear from (\ref{H_comp}) that it also has the
correct limits as $h/a\rightarrow 0$ and
$h/a\rightarrow\infty$. However,
it is also evident from \ref{Fig:H_comp}(d)
that the theory is unsatisfactory when
$a\kappa\ll 1$. We discuss this limit in the next section, where
we shall show that when $a\kappa\ll 1$ some of the charge cloud of ions
that neutralizes the surface charge on the cylindrical wall of the pore
spills out of the ends of the pore, where it is less effective at generating
electroosmotic flow. The scenario is shown schmatically in 
\ref{Fig:schematic_overspill}.

\section{3. Charge overspill from the ends of the pore, $a\kappa\ll 1$}

\subsection{3.1. Overspill of charge from the end of a semi-infinite pore}

We consider a cylindrical pore of radius $a$, with surface charge density
$\sigma_c$. When the Debye length $\kappa^{-1}\gg a$,  the
equilibrium potential $\phi_0$ (\ref{phi0_cylinder}) in an infinitely long cylinder can be expanded as
\begin{equation}
\phi_0=\phi_a
\left(1+\frac{(\kappa   r)^2}{4}+\cdots\right),
\end{equation}
where
\begin{equation}
\phi_a=\frac{\sigma_c}{\epsilon\kappa I_1(\kappa a)}
\approx\frac{2\sigma_c}{\epsilon\kappa(\kappa a)}.
\label{phi1_defn}
\end{equation}
Thus the equilibrium potential $\phi_0$ and charge density
$\rho_0=-\epsilon\kappa^2\phi_0$
within the charge cloud vary little over the
cross-section of the pore. On the other hand, if the cylinder is not
infinitely long and uniform, $\phi_0$ and $\rho_0$ vary
in the axial ($z$) direction with a length scale $\kappa^{-1}$.
We can therefore consider the equilibrium potential $\phi_0$ 
within the cylindrical pore to be a function only of
$z$ \cite{singer09}.

We first consider a semi-infinite, charged cylindrical pore going from $z=0$
to $z=\infty$. The equilibrium potential $\phi_0$
satisfies a one-dimensional Poisson-Boltzmann
equation,
\begin{equation}
\frac{\text{d}^2\phi_0}{\text{d}z^2}=-\kappa^2\phi_0.
\label{1D_PB_eqn}
\end{equation}
The solution that tends to the uniform potential $\phi_a$ within the
pore as $z\rightarrow\infty$ far from the pore end at $z=0$, is
\begin{equation}
\phi_0=\phi_a-A\exp(-\kappa z),
\label{phi_semi_infinite}
\end{equation}
for some unknown constant $A$. 
The charge density within the charge cloud
inside the pore is $-\epsilon\kappa^2\phi_0$, and when the cylindrical
pore is infinite (and hence uniform) the charge per unit length in the
charge cloud is $-\pi a^2\epsilon\kappa^2\phi_a=-2\pi\sigma_ca$, equal
and opposite to the charge per unit length on the pore walls. When the
pore is semi-infinite, with a non-uniform charge
cloud (\ref{phi_semi_infinite}), the
total charge that is lost from within the pore is
\begin{equation}
q_\text{lost}=\pi a^2\epsilon\kappa^2\int_0^\infty A\exp(-\kappa z)\,\text{d}z
=\pi a^2\kappa\epsilon A.
\end{equation}
At the end of the pore ($z=0$) the potential
is $\phi=\phi_a-A$.  

In $z<0$ the charge
cloud is no longer confined by the walls of the cylindrical pore
and spreads out radially: it is no longer possible to assume that
$\phi_0$ is a function of $z$ alone. 
We therefore need to solve the linearized Poisson-Boltzmann equation in the half-space $z<0$,
with $\phi_0=\phi_a-A$ over the region $z=0$, $r<a$ and
$\partial\phi_0/\partial z=0$ on $z=0$, $r>a$. At large distances from the
end of the pore, the potential decays as $\exp(-\kappa R)/R$,
where $R=(z^2+r^2)^{1/2}$ is a spherical polar coordinate,
but in the important region $R=O(a)$ the potential can be approximated by
the electrostatic potential corresponding to a solution of the Laplace
equation (i.e. $\kappa=0$). Hence, from (\ref{chi_m}),
\begin{equation}
\phi_0=(\phi_a-A)\frac{2}{\pi}\tan^{-1}\left(\frac{1}{\sinh\xi}\right).
\label{laplace_disk}
\end{equation}
To relate the potential (\ref{laplace_disk}) to the amount of charge
in the overspilling charge cloud (in $z<0$), we note that
the charge on one side of a charged disk at uniform potential
$(\phi_a-A)$ in unbounded space is 
$q=4a\epsilon(\phi_a-A)$.
Alternatively, one can argue that
far from the plane $z=0$, the spherical distance $R\approx a\cosh\xi$,
so that the potential (\ref{laplace_disk}) is approximately
\begin{equation}
\phi_0\approx (\phi_a-A)\frac{2a}{\pi R}.
\end{equation}
In a spherically symmetric geometry this field corresponds to the far
field around a point charge of magnitude $8a\epsilon(\phi_a-A)$,
and the total surface charge on one side of the disk is 
$q=4a\epsilon(\phi_a-A)$, in agreement with the charge obtained by considering
the capacitance of the disk. The charge in the overspill charge cloud in $z<0$
is equal and opposite to $q$, and is therefore
\begin{equation}
q_\text{overspill}=-4a\epsilon\phi_0(z=0)=-4a\epsilon(\phi_a-A).
\label{capacitance_disk}
\end{equation}
But the total charge in the overspill outside the end of the pore must be
equal and opposite to the charge that has been lost from within the pore.
Hence
\begin{equation}
4a\epsilon(\phi_a-A)=\pi a^2\kappa\epsilon A,
\end{equation}
so that
\begin{equation}
A=\frac{\phi_a}{1+\pi a\kappa/4},
\end{equation}
and the potential at the end of the pore is
\begin{equation}
\phi_a-A=\frac{\phi_a}{1+4/(\pi a\kappa)}.
\end{equation}
The charge that has been lost from the end of the
pore is equivalent to the charge usually found in a pore of length
\begin{equation}
h_\text{lost}=-\frac{q_\text{overspill}}{2\pi a\sigma_c}
=\frac{4}{4\kappa+\pi a\kappa^2}.
\label{h_lost_one_end}
\end{equation}

\subsection{3.2. Overspill from the two ends of a finite pore}

We can now perform the same analysis for a pore that
occupies the region $-h/2<z<h/2$. The equilibrium
potential within the pore has the form
\begin{equation}
\phi_0=\phi_a-B\cosh(\kappa z)+C,
\label{phi0_inside_two_ends}
\end{equation}
where we have chosen the solution that is symmetric about the centre of
the pore at $z=0$. The charge that has been lost from within the
pore is
\begin{equation}
q_\text{lost}=-\epsilon\kappa^2\pi a^2
\left(\int_{-h/2}^{h/2}(\phi_0-\phi_a)\,\text{d}z\right)
=\epsilon\pi a^2\kappa^2\left\lbrack
\frac{2B}{\kappa}\sinh\left(\frac{\kappa h}{2}\right)-Ch\right\rbrack.
\label{q_lost_inside_two_ends}
\end{equation}
The total flux of electric field through the two ends of the pore
is
\begin{equation}
2\pi a^2\left.\frac{\partial\phi_0}{\partial z}\right|_{z=h/2}
=-2\pi a^2\kappa B\sinh\left(\frac{\kappa h}{2}\right).
\label{flux_though_two_ends}
\end{equation}
Comparing (\ref{q_lost_inside_two_ends}) and
(\ref{flux_though_two_ends}) we conclude that $C=0$. The potential
over the ends of the pore is
\begin{equation}
\phi_0(h/2)=\phi_0(-h/2)=\phi_a-B\cosh(\kappa h/2).
\end{equation}
The total charge in the two overspill charge clouds is therefore,
by (\ref{capacitance_disk}),
\begin{equation}
q_\text{overspill}=-8a\epsilon\left\lbrack\phi_a
-B\cosh\left(\frac{\kappa h}{2}\right)\right\rbrack,
\label{q_overspill_two_ends}
\end{equation}
and this must be equal to the charge (\ref{q_lost_inside_two_ends})
lost from within the pore. Hence
\begin{equation}
B=\frac{4\phi_a}{4\cosh(\kappa h/2)+\pi a\kappa\sinh(\kappa h/2)}
\label{B}
\end{equation}
and
\begin{equation}
\phi_a-B\cosh\left(\frac{\kappa h}{2}\right)=
\frac{\phi_a\pi a\kappa\sinh(\kappa h/2)}{4\cosh(\kappa h/2)+\pi
  a\kappa\sinh(\kappa h/2)}.
\label{phi_a_minus_B_cosh}
\end{equation}
The total charge that has been lost (from the two ends) is equivalent to a
total lost length
\begin{eqnarray}
h_\text{lost}=-\frac{q_\text{overspill}}{2\pi a\sigma_c}&=&
\frac{8\sinh(\kappa h/2)}
{4\kappa\cosh(\kappa h/2)+\pi a\kappa^2\sinh(\kappa h/2)}
\label{h_lost_two_ends}
\\
&\sim&\frac{8}{4\kappa+\pi a\kappa^2},\hskip 30pt \kappa h\gg 1,
\label{h_lost_two_ends_h_large}
\\
&\sim&\frac{4h}{4+\pi ah\kappa^2/2}
,\hskip 30pt \kappa h\ll 1.
\label{h_lost_two_ends_h_small}
\end{eqnarray}
We see from eqs (\ref{h_lost_one_end}) and (\ref{h_lost_two_ends_h_large}) 
that when $\kappa h\gg 1$ the lost charge is twice that lost
from a single end of a pore. We also note that $h-h_\text{lost}>0$, and that
\begin{equation}
h-h_\text{lost}
\sim\frac{\pi a \kappa^2h^2}{8+\pi ah\kappa^2}
,\hskip 30pt \kappa h\ll 1.
\label{h_min_h_lost_h_small}
\end{equation}

\subsection{3.3. Overspill from the membrane surface into the pore}

If the cylindrical
pore itself is uncharged, but the membrane surfaces are charged, ions
from the charge cloud adjacent to the membrane surface
are able to move into the ends of the pore.

If the membrane has zero thickness,
the charge density $\rho_0$ in the equilibrium charge cloud is
given by (\ref{eq:rho0}), and
both $\rho_0$ and the potential $\phi_0=-\rho_0/(\epsilon\kappa^2)$
vary over the area of the pore. Nevertheless, we may work out the mean potential
over the circular pore,
\begin{equation}
\overline{\phi}_0=-\frac{1}{\epsilon\kappa^2\pi a^2}
\int_0^a 2\pi r\rho_0\,\text{d}r
=\frac{2\sigma_m}{\epsilon a}
\left\lbrack\frac{a}{2\kappa}-
 \int_0^\infty
\frac{aJ_1(as)J_1(as)}{s(\kappa^2+s^2)^{1/2}}\,\text{d}s
 \right\rbrack,
\end{equation}
where, when $a\kappa\ll 1$,
\begin{equation}
\int_0^\infty
\frac{aJ_1(as)J_1(as)}{s(\kappa^2+s^2)^{1/2}}\,\text{d}s
\approx a^2\int_0^\infty
\frac{J_1(t)J_1(t)}{t^2}\,\text{d}t
=\frac{4a^2}{3\pi}.
\end{equation}
Thus when the membrane has zero thickness (and there is no
cylindrical pore into which ions can escape)
the absence of surface charge over
the area of the pore
changes the average potential over the opening from 
the value
$\phi_0=\sigma_m/(\epsilon\kappa)$ due to a uniformly charged surface
to $\beta\sigma_m/(\epsilon\kappa)$, where
\begin{equation}
\beta\approx 1-\frac{8a\kappa}{3\pi},\hskip 20pt a\kappa \ll 1.
\label{beta_defn}
\end{equation}

We now consider the charge
that leaks into a pore of length $h>0$ from the charge clouds on either
side of the membrane.
We suppose that the potential on the planes
$z=\pm h/2$ is perturbed by an amount $D$, and becomes
\begin{equation}
\phi_0=\frac{\beta\sigma_m}{\epsilon\kappa}+D,\hskip 20pt z=\pm h/2.
\label{underspill_perturbed_potential}
\end{equation}
Within the pore, the potential obeys the 1-dimensional
Poisson-Boltzmann equation (\ref{1D_PB_eqn}), with solution
\begin{equation}
\phi_0=\left(\frac{\beta\sigma_m}{\epsilon\kappa}+D\right)
\frac{\cosh(\kappa z)}{\cosh(\kappa h/2)},
\end{equation}
and the additional charge within the pore is
\begin{equation}
q_\text{in}-\epsilon\kappa^2\pi a^2\int_{-h/2}^{h/2}\phi_0\,\text{d}z
=-\pi a^2\left(\beta\sigma_m+D\epsilon\kappa\right)
\frac{2\sinh(\kappa h/2)}{\cosh(\kappa h/2)}.
\end{equation}
Outside the pore, the perturbed potential
(\ref{underspill_perturbed_potential}) is associated with
a total additional charge (\ref{capacitance_disk})
\begin{equation}
q_\text{out}=-8a\epsilon D
\label{q_out_underspill}
\end{equation}
on the two sides of the membrane.
But the total change in charge caused by this redistribution must be zero,
i.e. $q_\text{in}+q_\text{out}=0$. Hence
\begin{equation}
\pi a^2\left(\beta\sigma_m+D\epsilon\kappa\right)
\frac{2\sinh(\kappa h/2)}{\cosh(\kappa h/2)}
+8a\epsilon D=0,
\end{equation}
i.e.
\begin{equation}
D=-\frac{\pi a\beta\sigma_m\sinh(\kappa h/2)}
{[4\cosh(\kappa h/2)+\pi a\kappa\sinh(\kappa h/2)]\epsilon}.
\end{equation}
The total charge $q_\text{in}=-q_\text{out}$ (\ref{q_out_underspill})
that leaks into the pore at the two ends
corresponds to the charge inside a uniformly charged
cylinder with surface charge density $\sigma_m$,  of length
\begin{eqnarray}
h_\text{gained}=-\frac{8a\epsilon D}{2\pi a\sigma_m}&=&\frac{4 a\beta\sinh(\kappa h/2)}
{4\cosh(\kappa h/2)+\pi a\kappa\sinh(\kappa h/2)}=\frac{a\kappa\beta}{2}h_\text{lost},
\label{h_gained}
\\
&\sim&\frac{a\kappa\beta h}{2},\hskip 20pt \kappa h\ll 1,
\label{h_gained_h_small}
\\
&\sim&\frac{a\beta}{2+\pi a\kappa},\hskip 20pt \kappa h\gg 1.
\end{eqnarray}
Thus $h_\text{gained}$ (\ref{h_gained}) is smaller than
$h_\text{lost}$ (\ref{h_lost_two_ends}) by a factor $a\kappa\beta/2$.
We can compare the
predictions (\ref{h_lost_two_ends})
and (\ref{h_gained}) against results
obtained from full numerical solution of the nonlinear Poisson-Boltzmann
equation with either $\sigma_m=0$ and
$ae\sigma_c/(\epsilon kT)=a\kappa e\zeta_c/(kT)=0.00273$,
or $\sigma_c=0$ and
$ae\sigma_m/(\epsilon kT)=0.00273$:
results for $a\kappa=0.1$ are given in Table 1. We see that there is
excellent agreement between the numerical computations and the analysis
presented above.

\begin{table}
\begin{center}
\begin{tabular}{cc @{\hspace{30pt}} cc @{\hspace{30pt}} cc}
$h/a$&$h\kappa$&\multicolumn{2}{c}{$h_\text{lost}/a$}
&\multicolumn{2}{c}{$h_\text{gained}/a$}
\\
&&theory (\ref{h_lost_two_ends})&numerical&theory (\ref{h_gained})&numerical
\\
10.0&1.0&8.9186&8.9249&0.4081&0.4119
\\
1.0&0.1&0.9953&0.9954&0.0455&0.0459
\\
0.1&0.01&0.1000&0.1000&0.0046&0.0046
\end{tabular}
\end{center}
\caption{The charge lost from the ends of a charged
pore when the membrane charge density $\sigma_m=0$, in terms of an
equivalent pore length $h_\text{lost}$ (\ref{h_lost_two_ends}),
and the charge gained inside an uncharged pore ($\sigma_c=0$)
from the charge cloud adjacent to the charged membrane surfaces,
in terms of an equivalent pore length $h_\text{gained}$ (\ref{h_gained}).
$a\kappa=0.1$.}
\end{table}

\subsection{3.4. Composite electroosmotic coefficient}

We first consider how the electroosmotic coefficients $H_c$ and $H_m$
are modified by the overspill of the charge cloud from inside the
cylindrical pore to outside the membrane.
If a uniform electric field of
strength $E=-\phi_1/h$ is applied between the ends of the pore,
the Navier Stokes equations for steady flow yield the axial velocity profile
\begin{equation}
u=\frac{a^2-r^2}{4\mu}\left(\rho_0E-\frac{{\rm d}p}{{\rm d}z}\right)
\end{equation}
so that the volumetric flow rate is
\begin{equation}
Q=\frac{a^4\pi}{8\mu}\left(\rho_0E-\frac{{\rm d}p}{{\rm d}z}\right).
\label{Q_in_uniform_pore}
\end{equation}
But $Q$ is independent of $z$ (by incompressibility), and the pressure $p$
difference between the two ends of the capillary is zero. Hence, integrating
(\ref{Q_in_uniform_pore}) along the length $h$ of the cylindrical pore,
and noting that
the total amount of charge in the charge cloud remaining within
the pore is
$2\pi a\sigma_c(h-h_\text{lost})$, we find
\begin{equation}
Q=\frac{a^4\pi E}{8h\mu}\int_{-h/2}^{h/2}\rho_0\,\text{d}z
=\frac{\pi a^3\sigma_c(h-h_\text{lost})\phi_1}{4h\mu}
=H_c\phi_1,
\label{H_c_underspill}
\end{equation}
which may be compared to the result (\ref{H_c_kappa_small})
which ignores overspill.
The charge cloud outside the pore is enhanced by the overspill, and becomes
(in $z>0$)
\begin{equation}
\rho_0
=-\sigma_m\kappa\exp(-\kappa z)
-\frac{2\epsilon\kappa^2}{\pi}\left\lbrack\phi_a-B\cosh(\kappa h/2)\right\rbrack
\tan^{-1}\left(\frac{1}{\sinh\xi}\right),
\end{equation}
with the final term $[\phi_a-B\cosh(\kappa h/2)]$,
corresponding to the overspill charge cloud
(\ref{phi_a_minus_B_cosh}), being
approximately valid in a volume $O(a^3)$ around the pore, but invalid at large
distance $O(\kappa^{-1})$ from the pore, where the exponential decay
of the charge density
is not captured by the solution (\ref{laplace_disk}) of the Laplace equation.
The volumetric flow rate through a pore of zero thickness created by
a potential difference $\phi_1$
is given by the integral (\ref{reciprocal_integral})
and was shown by Mao et al. \cite{mao14} to be:
\begin{eqnarray}
Q  &=&  \frac{2a^3\phi_1}{\pi\mu}
\int_0^{\frac{\pi}{2}} d\eta \int_0^\infty \rho_0
\frac{\cos^2\eta\sin\eta}{\cosh\xi} d\xi
\nonumber\\
&=&-\frac{a^3\kappa\sigma_m\phi_1}{3\mu}
-\frac{4\epsilon\kappa^2a^3\phi_1}{\pi^2\mu}(\phi_a-B)
\int_0^{\pi/2}\cos^2\eta\sin\eta\text{ d}\eta \int_0^\infty 
\tan^{-1}\left(\frac{1}{\sinh\xi}\right)\frac{\text{d}\xi}{\cosh\xi}
\nonumber\\
&=&-\frac{a^3\kappa\sigma_m}{3\mu}\phi_1
-\frac{4\epsilon a^3\kappa^2}{3\pi^2\mu}\phi_1(\phi_a-B)I_3,
\label{Q_integral_disk}
\end{eqnarray}
where
\begin{equation}
I_3=
\int_0^\infty 
\tan^{-1}\left(\frac{1}{\sinh\xi}\right)\frac{\text{d}\xi}{\cosh\xi}
=\int_0^\infty \tan^{-1}x\ \frac{\text{d}x}{1+x^2}
=\frac{\pi^2}{8}.
\end{equation}
Hence the electroosmotic flow rate $Q=H_m\phi_1$
due to the charge cloud outside the
membrane is modified, and $H_m$ becomes
\begin{equation}
H_m=\frac{a^3\kappa}{3\mu}\left\lbrack
\sigma_m+
\frac{\pi\sinh(\kappa h/2)\sigma_c}
{4\cosh(\kappa h/2)+\pi a\kappa\sinh(\kappa h/2)}
\right\rbrack.
\label{H_m_overspill}
\end{equation}

If $\sigma_m$ is comparable to $\sigma_c$, then we saw in Section 3.3 that
the change in the charge within the pore due to the charge cloud outside
the membrane entering the pore is $O(a\kappa)$ smaller than the loss of
charge from the charge cloud within the pore to the regions
outside the membrane. However, this contribution can be included
with very little effort, and becomes important in the limit
$h\rightarrow 0$, when the gain (\ref{h_gained_h_small})
in charge within the pore
from the outside surface charge density $\sigma_m$
is proportional to $h_\text{gained}\propto h$,
whereas the charge cloud (due to $\sigma_c$ within the pore)
remaining within the pore, is
proportional to $h-h_\text{lost}\propto h^2$, by
(\ref{h_min_h_lost_h_small}).
The electroosmotic coefficient $H_c$
for the cylindrical pore (\ref{H_c_underspill}) becomes
\begin{equation}
H_c=\frac{\pi a^3\sigma_c(h-h_\text{lost}+h_\text{gained}\sigma_m/\sigma_c)}
{4h^2\mu},
\label{H_c_overspill_underspill}
\end{equation}
and the electroosmotic coefficient $H_m$ for the charge cloud
outside the membrane (\ref{H_m_overspill}) becomes
\begin{equation}
H_m=\frac{a^3\kappa}{3\mu}\left\lbrack
\sigma_m+
\frac{\pi\sinh(\kappa h/2)(\sigma_c-a\kappa\beta\sigma_m/2)}
{4\cosh(\kappa h/2)+\pi a\kappa\sinh(\kappa h/2)}
\right\rbrack.
\label{H_m_overspill_underspill}
\end{equation}

\begin{figure}
\begin{center}
\includegraphics[width=0.7\textwidth]{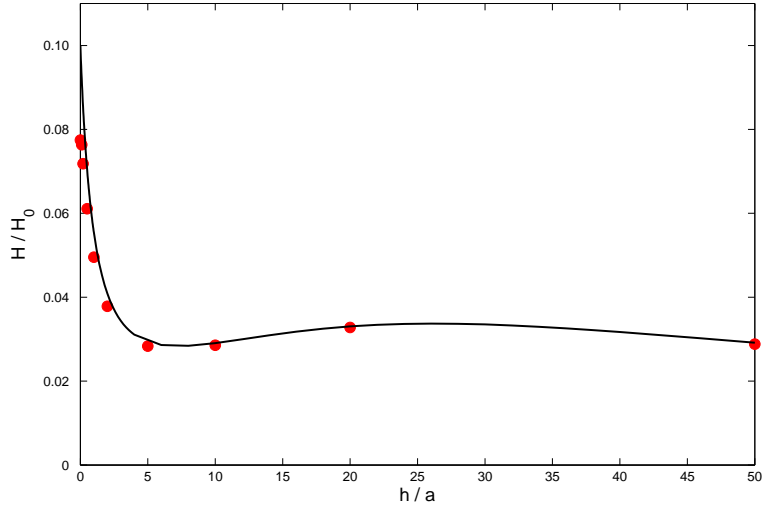}
\caption{\label{Fig:H_comp_akappa_small}
The electroosmotic coefficient $H$ scaled by $H_0$ (\ref{H0_defn}) for
$\sigma_m=\sigma_c$, as a function of $h/a$, for
$a\kappa=0.1$, including the effect of overspilling charge clouds.
\copy\grapha $H_\text{comp}$ (\ref{H_comp}),
using $H_c$ given by (\ref{H_c_overspill_underspill}) and $H_m$ given by (\ref{H_m_overspill_underspill});
solid circles: full PNP numerical computation. c.f. Fig. \ref{Fig:H_comp}(d), in
which overspill was neglected.}
\end{center}
\end{figure}

Now that $H_c$ (\ref{H_c_overspill_underspill}) and 
$H_m$ (\ref{H_m_overspill_underspill})
have been corrected for the effects of overspill
in the two directions, they can be inserted into the expression
(\ref{H_comp})
for the composite electroosmotic coefficient $H_\text{comp}$.
Results are shown in \ref{Fig:H_comp_akappa_small}, together
with full numerical solutions of the Poisson-Nernst-Planck equations.
We see that the agreement between theory and computation is much better
than  when overspill is ignored (\ref{Fig:H_comp}(d)). 
Charge overspill or underspill causes the total charge of mobile ions within the 
pore to be different from what might be expected on the basis of net 
electroneutrality of the pore. 
Thus, the driving force is modified leading to deviations from the calculated 
result that ignores such effects. The ``lost length'' $h_\text{lost}$ in (\ref{H_c_overspill_underspill})
restores  this effect.

Note that when $h\ll\kappa^{-1}$ the effective length of the
cylindrical pore $h-h_\text{lost}\approx \pi a\kappa^2h/2$,
by (\ref{h_min_h_lost_h_small}). The approximation
(\ref{H_c_overspill_underspill}) for $H_c$
is therefore dominated by the term $h_\text{gained}$,
and gives $H_c\sim \pi a^4\kappa\beta\sigma_m/(8h\mu)$, with
$H_c/H_m\sim 3\pi a\beta/(8h)$. We conclude from
(\ref{H_c_for_H_comp_increasing}) that $H_\text{comp}$
is a decreasing function of $h$ near $h=0$, as seen in 
\ref{Fig:H_comp_akappa_small}.

\section{4. Numerical simulation}

We solve the time--independent PNP--Stokes equations numerically using the finite volume method. The governing equations are:
\begin{eqnarray}
\epsilon \nabla^2 \phi + \sum_{i=1}^{N} z_i e n^i & = & 0,
\label{eq:poisson}
\\
\nabla\cdot\left\lbrack n^i\mathbf{u} -\omega^i(kT \nabla
n^i + ez_in^i\nabla\phi) \right\rbrack&=&0 ,
\label{eq:NP}
\\ 
-\nabla p + \mu \nabla^2 \mathbf{u} -  \nabla \phi \sum_{i=1}^{N} z_ien^i & = & 0, \label{eq:stokes}\\
\nabla \cdot \mathbf{u} & = & 0. \label{eq:continuity}
\end{eqnarray}
Here $n^i$ is the local concentration of the i--th ion, $i=1 \sim N$ and $N=2$ in this case, $z_i$ is the valence of the i--th ion, and in our case we have $z_1 = -z_2 = 1$, $\omega^i$ is the mobility of the i--th species of ion. The electric potential is $\phi$, the flow velocity is $\mathbf{u}$ and $p$ is the pressure.

We solve the system of coupled equations (\ref{eq:poisson})-(\ref{eq:continuity}) 
in the axisymmetric geometry depicted in \ref{Fig:schematic}. Thus, we have a cylindrical pore of radius $a$ and length $h$ connecting two large cylindrical reservoirs of radius $b$. The lengths of AB and EF in our simulation are also taken to be $b$. Numerically, $b$ is kept much larger than either $a$ or the Debye length 
$\kappa^{-1}$ so that the reservoirs are effectively infinite.

We apply the following boundary conditions. At A and F, the two ends of the reservoirs, ion concentrations are made equal to the concentration in the bulk electrolyte, or $n^i = n^{i}_{\infty}$; a potential difference of $\Delta V$ is applied across the system by setting $\phi$ to $\pm \Delta V/2$ respectively at A and F while the pressure at these two ends are set equal to the bulk pressure, $p = p_\infty$. At AB and EF, the side walls of the cylindrical reservoir, the radial component of the electric field, ionic flux and velocity are all set to zero, as the walls are considered to be far away from the pore. A zero tangential shear stress is imposed on AB and EF as well. At the membrane and pore surfaces BC, CD and DE, a no--flux condition is used for (\ref{eq:NP}), and a no--slip condition is used  for the flow. At solid fluid interfaces 
the electric potential is continuous but the normal component of the electric field undergoes a jump 
$[ \epsilon \mathbf{E}\cdot \hat{n} ] = \sigma_j$, where $j = m$ at BC and DE, $j=c$ at CD
and we have used square brackets to indicate change of the enclosed quantity across the interface.
The suffixes $m$ and $c$ indicate the surface charge densities on the face of the membrane and within the pore respectively.  Here $\hat{n}$ is the unit normal at the surface directed into the fluid.

An electrohydrodynamic solver was implemented to solve the system described above using the OpenFOAM CFD library \cite{OPENFOAM}, a C++ library designed for computational mechanics. Structured mesh is constructed with polyMesh, a meshing tool built in OpenFOAM. The grid is refined near the membrane and pore surfaces to resolve the Debye layer. Grid independence was checked in all cases by refining the grid and verifying that the 
solution does not change within specified tolerances.

For the finite volume discretization of the governing equations, central differences are used for all diffusive terms in (\ref{eq:NP}) and viscous terms in (\ref{eq:stokes}). A second--order upwind scheme is used for the convective terms in (\ref{eq:NP}). The discretized linear system is solved using a pre-conditioned conjugate gradient solver if the matrix is symmetric or a pre-conditioned bi--conjugate gradient solver if the matrix is 
asymmetric \cite{ferziger&peric}. 

An iterative scheme is used to solve the PNP--Stokes equations. Initially, the flow velocity 
is set to zero. Equations  (\ref{eq:poisson}) and (\ref{eq:NP}) are then solved sequentially in a loop with under-relaxation (to ensure stability of the nonlinear PNP system) until the absolute residual is smaller than a specified tolerance -- in our case, $10^{-6}$.
 The electric force density $- \nabla \phi\sum_i z_ien^i $ is then obtained from this solution and used 
 as an explicit external forcing in the next step - the solution of the incompressible Stokes flow problem:
 (\ref{eq:stokes}) and (\ref{eq:continuity}). The SIMPLE algorithm is used to solve the hydrodynamic problem.  The flow field so computed is then substituted into (\ref{eq:NP}) and the PNP equations are then solved again using the updated flow field. An outer loop is constructed to iterate over the PNP loop and Stokes flow module, until the solution changes negligibly between two outer iterations.

Our main object of interest is the volumetric flux, $Q$. This is obtained in a post processing 
step by numerically integrating the axial velocity over the plane $z = 0$.
At the low voltages employed, a substantially linear relation is found between 
$Q$ and $\Delta V$, the electroosmotic coefficient $H$ is then obtained from $H = Q/ \Delta V$.
The quantities $h_{lost}$ and $h_{gained}$ were obtained numerically by integrating the local charge density 
within the confines of the pore.

\section{4. Concluding remarks}

The analysis presented here shows that it is possible to use simple
analyses based on continuity of volumetric flow rate and electric
current to estimate electroosmotic
end effects in a charged cylindrical pore traversing
a membrane of thickness $h>0$. Note that we have made repeated
use of the assumption that surface charge densities, and corresponding
zeta potentials, are small. Not only have we worked
with the linearized Poisson-Boltzmann equation
(\ref{linear_poisson_boltzmann_eqn}), but we have used superposition to
combine various contributions to the charge clouds due to
overspill of the clouds from one region (inside/outside the pore) to the other.
At high potentials it would also be necessary to keep track of the
fluxes of individual ion species, rather than simply ensuring
that the total electrical current is continuous \cite{biscombe12}.

The assumption of small potentials also justifies the neglect of other 
nonlinear electrokinetic effects such as Induced Charge
Electroosmosis (ICEO) \cite{murtsovkin96, Squires2004} which could 
produce vortices in the vicinity of sharp corners \cite{Thamida2002}
or near rapid constrictions in channels \cite{park_eddies_2006}.
However, numerical solutions confirm the expectation that the flow rate is only weakly affected by 
such vortices, particularly under conditions of small potentials \cite{Mao2013}

Recent experimental work
\cite{ghosal2013Nanoletter,Keyser2006,Garaj2010,Schneider2010,Merchant2010} 
on nanopores report potential differences of $\Delta\phi \sim 0 - 200$~mV applied across the pore.
Here we have assumed that $\Delta\phi\ll\zeta$, where 
$\zeta$ itself is assumed small in comparison with the thermal voltage $kT/e \sim 25$ mV.
 Thus, our results can only be expected to describe the initial linear part of the current-voltage 
 and flow-voltage characteristics, even though numerical simulations seem to show \cite{Mao2013}
 that this linear regime extends to applied voltages $\sim 100$mV.

Finally, we point out that the
correction factor $\beta$ (\ref{beta_defn})
reminds us that the hole in the charged membrane removes a circular
region of surface charge and reduces the equilibrium potential
at the entrance to the pore. The introduction of
$\beta<1$ improved the agreement between
theoretical and numerical results for $h_\text{gained}$ in Table 1.
However, the analysis is not rigorous, since the equilibrium
potential across the hole is not uniform. The $O(1-\beta)$
correction to the equilibrium potential corresponds to an $O(1-\beta)$
correction
to the charge density $\rho_0$.
If we use this in the integral expression
(\ref{Q_integral_disk})
in order to determine a correction to the electroosmotic flow rate
through a membrane of zero thickness, the analysis suggests that the correction
to the leading order result (\ref{H_m_akappa_small})
for $a\kappa\ll 1$ should be $O((a\kappa)^2)$,
whereas investigation of the difference (seen in \ref{Fig:H_m_log_log})
between numerical results and the asymptote (\ref{H_m_akappa_small})
indicates additional
corrections $O((a\kappa)^2\ln a\kappa)$.

\begin{acknowledgement}
JDS thanks the Department of Applied Mathematics, University of Cambridge, and
the Institut de M\'ecanique des Fluides de Toulouse, for hospitality.
MM \& SG acknowledge support from the NIH through Grant 4R01HG004842.
\end{acknowledgement}

\newpage

\end{document}